\documentclass[showpacs,twocolumn,prb,preprintnumbers,amsmath,amssymb]{revtex4}
\usepackage[latin5]{inputenc}
\usepackage{graphicx}
\usepackage{dcolumn}
\usepackage{bm}

\begin{document}

\title{Auger recombination and carrier multiplication in embedded silicon and 
germanium nanocrystals}

\author{C. Sevik}
\author{C. Bulutay}
\affiliation{Department of Physics, Bilkent University, Bilkent, Ankara, 06800, Turkey}
\date{\today}
\begin{abstract}
For Si and Ge nanocrystals (NCs) embedded in wide band-gap matrices, 
Auger recombination (AR) and  carrier multiplication (CM) lifetimes are computed 
exactly in a three-dimensional real space grid using empirical 
pseudopotential wave functions. Our results in support of recent experimental data offer new 
predictions. We extract simple Auger constants valid for NCs. We show that both Si and Ge NCs can
benefit from photovoltaic efficiency improvement via CM due to the fact that under an optical 
excitation exceeding twice the band gap energy, the electrons gain lion's share from the 
total excess energy and can cause a CM. We predict that CM becomes especially efficient for 
hot electrons with an excess energy of about 1~eV above the CM threshold.
\end{abstract}



\maketitle

\section{Introduction}
The Group IV semiconductor, Si and to a lesser extend, Ge have been indispensable for the 
electronics and photovoltaics industry. The recent research efforts have shown that their 
nanocrystals (NCs) bring new features which fortify their stands. For 
instance, NCs can turn these indirect band-gap bulk materials into light 
emitters~\cite{ossicini} or offer increased efficiencies in solar cells.~\cite{nozik-jap}
The latter has been demonstrated in a very recent experimental study~\cite{nozik-nl} 
by significantly increasing the solar cell efficiency in colloidal Si NCs due to carrier 
multiplication (CM) which enables multiple exciton generation in response to a 
single absorbed photon.~\cite{schaller,ellingson}
Similarly, the inverse process, Auger recombination (AR) is also operational 
and it introduces a competing mechanism to CM which can potentially diminish the solar cell 
efficiency and in the case of light sources it degrades the performance by inflating the 
nonradiative carrier relaxation rate.~\cite{klimov-rev}

The utilization and full control of both CM and AR require a rigorous 
theoretical understanding. The pioneering series of theory publications on the AR in 
Si NCs belong to a single group based on an atomistic tight binding approach.~\cite{mihalcescu,delerue1,delerue2}
Unfortunately, they only considered hydrogen passivated Si NCs without addressing the shape 
effects. Moreover, their results do not reveal a size-scaling for AR but rather a 
scattered behavior over a wide band of lifetimes in the range from few picoseconds to few nanoseconds 
as the NC diameter changes from 2 to 4~nm. In the past decade, no further theoretical assessment of AR 
in Si NCs was put forward. In this context, the Ge NCs have not received any attention although 
with their narrower effective band gap, they can benefit more from the low-energy part of 
the solar spectrum in conjunction with CM for increasing the efficiency.

\begin{figure}[h!]
\includegraphics[width=6cm]{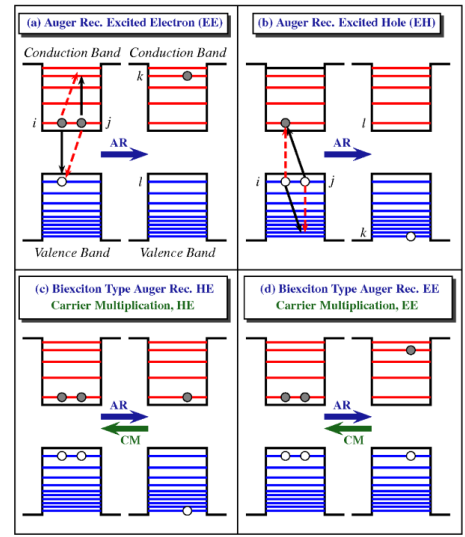}
\caption{\label{AugerRec}(Color online) AR in NCs: (a)~excited electron,
(b)~excited hole; the solid and dashed arrows refer to direct and exchange processes. 
Biexciton type AR and its inverse process CM: (c)~excited electron (d)~excited hole.}
\end{figure}

In this work we provide a theoretical account of AR and CM in Si and Ge NCs 
which reveals their size, shape and energy dependence. Another important feature of this work, unlike 
commonly studied hydrogen-passivated NCs is that we consider NCs \textit{embedded} in a wide band-gap 
matrix which is essential for the solid-state device realizations.
Similar to the classification of Wang \emph{et al.} in their theoretical 
work on Coulombic excitations in CdSe NCs,~\cite{wang03} we consider different possibilities of AR  
as shown in Fig.~\ref{AugerRec}. We use the type of the exicted carrier as the discriminating label, 
hence we have the excited electron (Fig.~\ref{AugerRec}(a)) and the excited hole 
(Fig.~\ref{AugerRec}(b)) AR and their biexciton variants (Fig.~\ref{AugerRec}(c) and (d)).
All of these have corresponding CM processes taking place in the reverse direction but only the CM in 
Fig.~\ref{AugerRec}(c) and (d) are studied as they can be optically induced.

\begin{figure}[h!]
\includegraphics[width=8cm]{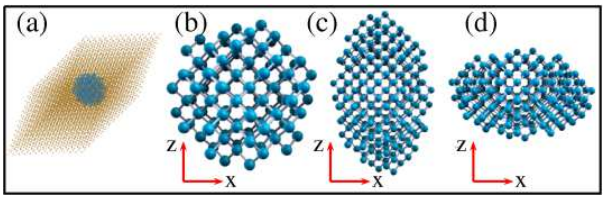}
\caption{\label{shapeNC}(Color online) (a) Embedded NC in a supercell, core atoms of a (b) spherical, 
(c) oblate and (d) prolate ellipsoidal NC.}
\end{figure}

\section{Theoretical Details}
Both AR and CM require an accurate electronic structure over a wide energy band extending up to at 
least 3-4~eV below (above) the highest occupied molecular orbital-HOMO (lowest unoccupied 
molecular orbital-LUMO). Another constraint is to incorporate several thousands of core 
and host matrix atoms within a supercell (see Fig.~\ref{shapeNC} (a)). To meet these 
requirements we have employed the linear combination of bulk bands basis within the 
empirical pseudopotential framework which can 
handle thousands-of-atom systems both with sufficient accuracy and efficiency over a 
large energy window.~\cite{wang97}  Details regarding its performance and the 
implementation such as the wide band-gap host matrix can be found in Ref.~\onlinecite{bulutay07}. 
We should mention that Califano \emph{et al.} have successfully employed a very similar 
theoretical approach in order to explain the hole relaxation in CdSe NCs.~\cite{califano}

After solving the atomistic empirical pseudopotential Hamiltonian for the energy levels 
and the wave functions, the AR and CM probability can be extracted using the Fermi's golden rule,
\begin{equation}
R = \frac{\Gamma}{\hbar}\displaystyle\sum_{f}
\frac{\left|\left\langle\psi_{i}\left|V_{c}(\mathbf{r_{1}},\mathbf{r_{2}})\right|
\psi_{f}\right\rangle\right|^2}{(E_{f}-E_{i})^{2}+(\Gamma/2)^{2}},\label{fermi}
\end{equation}
where $\psi_{i}$ and $\psi_{f}$ are respective initial and final configurations with the corresponding energies 
$E_{i}$ and $E_{f}$, respectively, and $\Gamma$ is the level broadening parameter which is taken as 
10~meV. However sensitivity to this parameter is also considered in this work. 
The spin-conserving screened Coulomb potential is given by
$ V_{c}(\mathbf{r_{1}},\mathbf{r_{2}})=e^{2}/\epsilon(\mathbf{r_{1}},
\mathbf{r_{2}})|\mathbf{r_{1}}-\mathbf{r_{2}}|$.
The subject of screened Coulomb interaction in NCs is an active source of debate; recent publications
predict reduced screening~\cite{ogut03,trani06} whereas, other theoretical investigations~\cite{allan,wang03} 
have concluded that the inverse dielectric function is 
bulklike inside the NC. Therefore, we follow~\cite{wang03} and use
\begin{eqnarray}
\frac{1}{\epsilon(\mathbf{r_{1}},\mathbf{r_{2}})}=\frac{1}{\epsilon_{\begin{tiny}\mbox{out}\end{tiny}}}+
\left(\frac{1}{\epsilon_{\begin{tiny}\mbox{in}\end{tiny}}}-
\frac{1}{\epsilon_{\begin{tiny}\mbox{out}\end{tiny}}}\right)m(\mathbf{r_{1}})m(\mathbf{r_{2}}),\label{dielec}
\end{eqnarray}
as the inverse dielectric function, where, the so-called mask function $m(\mathbf{r})$ is set to 
1 when $\mathbf{r}$ inside of the NC and 0 when $\mathbf{r}$ outside of the NC.

Expressing the initial and final states of the AR shown in Fig.~\ref{AugerRec} (a) or (b) by using the Slater 
determinant, the matrix elements ($\left\langle \psi_{i}\left|V_{c}(|\mathbf{r_{1}},\mathbf{r_{2}}|)
\right|\psi_{f}\right\rangle$) can be calculated as
\begin{eqnarray}
M(i,j;k,l) & = &\frac{1}{V^{2}}\int \int \phi_{k}^{*}(\mathbf{r_{1}})\phi_{l}^{*}(\mathbf{r_{2}})V_{c}(\mathbf{r_{1}},
\mathbf{r_{2}})\nonumber \\ 
\hspace{-0.8cm}& & \times (\phi_{i}(\mathbf{r_{1}}) \phi_{j}(\mathbf{r_{2}})-\phi_{j}(\mathbf{r_{2}}) 
\phi_{i}(\mathbf{r_{1}}))d^{3}r_{1}d^{3}r_{2},
\end{eqnarray}
here the labels $i$, $j$ and $k$ and $l$ refer respectively to the initial and final states which also include 
the spin and $V$ is volume of the supercell.

\begin{figure}[h!]
\includegraphics[width=8cm]{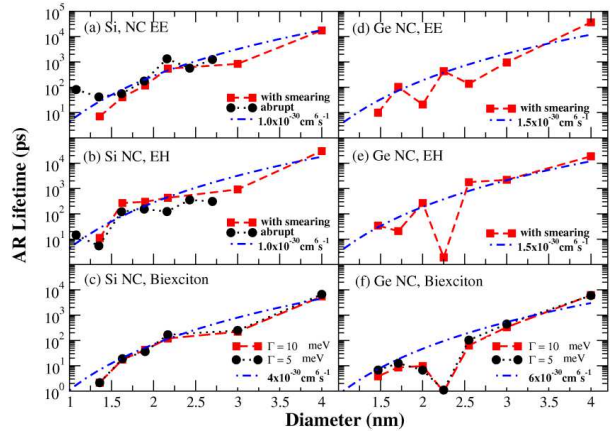}
\caption{\label{AugerRes}(Color online) AR lifetimes for (a)~excited electron, (b)~excited hole, and (c)~biexciton 
types in Si NCs, and (d)~excited electron, (e)~excited hole, and (f)~biexciton types in Ge NCs. 
Square symbols represent AR lifetimes with interface smearing, and dashed lines show AR lifetimes 
calculated from our proposed $C$ values. Spherical symbols in (a) and (b) represent AR lifetimes in Si NCs with abrupt interfaces.}
\end{figure}

These matrix elements, $M(i,j;k,l)$ are computed exactly in a three-dimensional real space grid
without resorting to any envelope approximation. 
The number of final states are determined setting the final state window to $\pm 7 \Gamma$ around 
the exact conserved energy $E_{k} (= E_{j}+E_{i}-E_{l})$.
For the initial states $i$ and $l$, Boltzmann average is taken into account due to thermal 
excitations. The same should apply to the other initial state $j$, however, as a safe but computationally 
very rewarding simplification it is kept fixed at LUMO for the excited electron (EE), and at HOMO 
for the excited hole (EH) type AR.

\section{Results and Discussions}
We first apply this formalism to spherical NCs (see Fig.~\ref{shapeNC}(b)) having abrupt interfaces. 
The corresponding AR lifetimes for EE and EH processes are plotted as a function of NC diameter 
in Fig.~\ref{AugerRes} (a) and (b). 
The $C_{3v}$ point symmetry of the NCs in the case of abrupt interface between NC core and the matrix causes 
oscillations in the physical quantities such as the state splittings and the density of states with respect 
to NC diameter.~\cite{bulutay07} When we account for the interface transition region between the NC and host 
matrix,~\cite{daldosso} we observe that these strong 
oscillations in the size dependence of AR are highly reduced for Si and Ge NCs (cf. Fig.~\ref{AugerRes}). 
The interface region especially 
affects the excited state wave functions and the final state density of states and it makes our model 
more realistic for both Si and Ge NCs. As an observation of practical importance, we can reproduce our 
data remarkably well (cf.~Fig.~\ref{AugerRes}) using  the simple expression $1/\tau=Cn^{2}$, 
with an Auger coefficient of $C=1\times 10^{-30}$ cm$^{6}$s$^{-1}$ for Si NCs and 
$C=1.5\times 10^{-30}$ cm$^{6}$s$^{-1}$ for Ge NCs, where  $n=2/V_{\begin{tiny}\mbox{NC}\end{tiny}}$
is the carrier density within the NC such that there should be two electrons or holes 
to initiate an AR.

The other two types of AR shown in Fig.~\ref{AugerRes} (c) and (f) refer to biexciton 
recombinations. This process becomes particularly important under high carrier densities 
such as in NC lasers or in solar cells under concentrated sunlight. Its probability can 
be expressed in terms of EE and EH type AR as,~\cite{wang03}
$1/\tau_{\begin{tiny}\mbox{XX}\end{tiny}}=2/\tau_{\begin{tiny}\mbox{EE}\end{tiny}}+2/\tau_{\begin{tiny}\mbox{EH}\end{tiny}}$
where $\tau_{\begin{tiny}\mbox{EE}\end{tiny}}$ and $\tau_{\begin{tiny}\mbox{EH}\end{tiny}}$ 
are EE and EH lifetimes. Fig.~\ref{AugerRes} (c) and (f) compares the 
computed biexciton type AR for Si and Ge NCs with the expression $1/\tau=Cn^{2}$ where the value 
$C=4\times 10^{-30}$cm$^{6}$s$^{-1}$ and $6\times 10^{-30}$cm$^{6}$s$^{-1}$ are used which are obtained from the previous 
$C$ values extracted for EE and EH processes together with the $\tau_{\begin{tiny}\mbox{XX}\end{tiny}}$ expression. 
For Si NC case, our calculated value at 3~nm diameter agrees reasonably well with the experimental photoluminescense decay time of about 105~ps which was attributed 
to AR.~\cite{troje} In Fig.~\ref{AugerRes} (c) and (f), we also demonstrate the fact that a choice of 
$\Gamma = 5$~meV does not introduce any marked deviation from the case of $\Gamma = 10$~meV 
as used in this work for both Si and Ge NCs. This parameter test automatically checks the sensitivity to the 
final state energy window chosen as $\pm$7$\Gamma$.

\begin{table}[h!]
\caption{\label{table1}AR lifetimes for different ellipsoidal shapes of Si NCs with diameters of 
1.63 and 2.16~nm.}
\begin{ruledtabular}
\begin{tabular}{lcccccc}
&\multicolumn{2}{c}{Spherical}&\multicolumn{2}{c}{Prolate}&\multicolumn{2}{c}{Oblate} \\
\cline{2-3} \cline{4-5} \cline{6-7}
$D$ (nm)& 1.63 & 2.16 & 1.63 & 2.16 & 1.63 & 2.16 \\\hline
EE (ps) & 40 & 541 & 32 & 103 & 36 & 121\\
EH (ps) &267 & 430 & 74 & 76 & 26 & 139\\
\end{tabular}
\end{ruledtabular}
\end{table}

Next, we demonstrate the effects of deviation from sphericity on Si NCs. We consider 
both oblate (Fig.~\ref{shapeNC}(c)) and prolate (Fig.~\ref{shapeNC}(d)) ellipsoidal Si NCs 
described by the ellipticity value of $e$=0.85.
For the comparison purposes, we preserve the same number of atoms used in spherical NCs of 
diameters 1.63 and 2.16~nm. The results listed in Table~\ref{table1} indicate that the spherical 
NC has a lower Auger rate than the aspherical shapes. This can be reconciled as follows: in 
the case of either prolate or oblate NC, the electronic 
structure is modified in such a way that the number of final states is increased, furthermore, 
a coalescence of the states around the HOMO and LUMO occurs. 
A similar effect was also observed in the asphericity-induced enhancement of Auger 
thermalization of holes in CdSe NCs.~\cite{califano}
However, we should note that the shape effects are not pronounced.

\begin{figure}[h!]
\includegraphics[width=8cm]{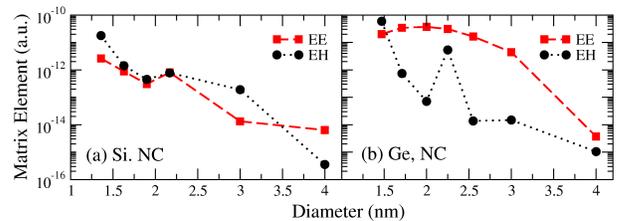}
\caption{\label{matel}(Color online) Average Coulomb matrix elements for (a) Si and (b) Ge NCs 
for EE type AR (red squares) and for EH type AR (black spheres).}
\end{figure}

In their work on the CM in PbSe NCs, Allan and Delerue have deduced that such 
Coulombic interactions are primarily governed by the state-density function.~\cite{allan06} 
Even though we agree on the importance of the density of states, we would like to emphasize 
the significant role of the Coulomb matrix elements. We illustrate our point by Fig.~\ref{matel}, where 
the average matrix element for Si and Ge NCs are shown. The strong size dependence leading 
to a variation over several orders of magnitude indicates their nontrivial role.

\begin{figure}[h!]
\includegraphics[width=8cm]{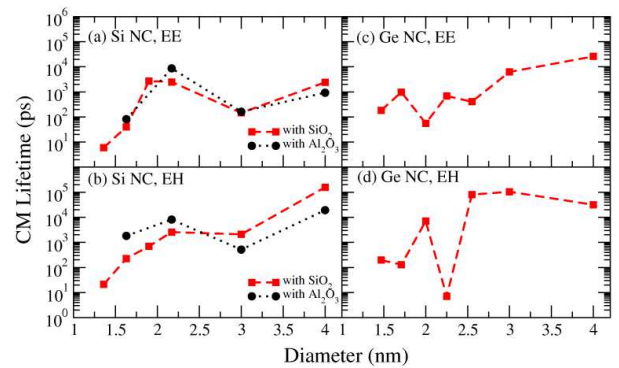}
\caption{\label{Si-Ge-Multip}(Color online) CM Lifetime results for (a)~EE and (b)~EH types in Si NCs 
embedded in SiO$_{2}$ and Al$_{2}$O$_{3}$, and (c)~EE and (d)~EH types in Ge NCs embedded in Al$_{2}$O$_{3}$.}
\end{figure}

Regarding the CM, we first consider the inverse Auger process (cf., Fig.~\ref{AugerRec} (c) and (d)) 
for different diameters of the Si and Ge NCs.
Therefore, we consider the impacting electron (hole) to have an energy of 
$E_{\mbox{\begin{scriptsize}gap\end{scriptsize}}}=
E_{\mbox{\begin{scriptsize}LUMO\end{scriptsize}}}-
E_{\mbox{\begin{scriptsize}HOMO\end{scriptsize}}}$ above (below) the conduction (valence) 
band edge i.e., just at the threshold energy to initiate a CM event. 
As seen in Fig.~\ref{Si-Ge-Multip}, EE and EH type CM lifetimes for Si and Ge 
NCs decrease from the few ns to few ps as the NC diameter decreases. 
However, for EE (EH) type CM, the small number of final states at the bottom of 
the CB (top of the VB) cause a nonmonotonic dependence of CM on size of the NC.

\begin{figure}[h!]
\includegraphics[width=8cm]{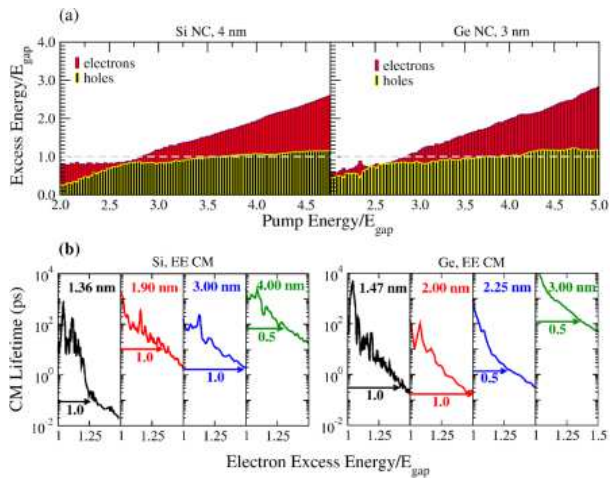}
\caption{\label{fig6}(Color online) (a) Electron and hole excess energy vs optical pump (excitation) 
energy for 4~nm Si and 3~nm Ge NCs, (b) CM lifetime vs electron excess energy for 
different diameter of Si and Ge NCs. The horizontal arrows provide the 1 and 0.5~eV marks.}
\end{figure}

From the practical point of view, the investigation of the effect of excess energy on the CM under 
an optical excitation above the effective gap, $E_{\mbox{\begin{scriptsize}gap\end{scriptsize}}}$ 
is even more important.
Here, the correct placement of excited electron and hole after optical excitation is critical.
We assign the excited electron and hole to their final states by accounting for all interband transitions 
with the given energy difference as 
weighted by the dipole oscillator strength of each transition which is a direct measure of the 
probability of that particular event.
In Fig.~\ref{fig6}~(a) we observe that the electrons receive 
the lion's share of the total excess energy which is the desired case for the high efficiency utilization 
of CM in photovoltaic applications.~\cite{werner}  Our threshold value of 
2.8~$E_{\mbox{\begin{scriptsize}gap\end{scriptsize}}}$ for Si NCs is somewhat
higher than the recent experimental data.~\cite{nozik-nl} 
In Fig.~\ref{fig6}~(b) we show the corresponding electron-initiated CM lifetimes as a function 
of excess energy. It can be observed that CM is enhanced by more than two orders of magnitude within an 
excess energy of 1~eV beyond the CM threshold reaching a lifetime of few picoseconds.

\section{Conclusions}
To summarize, we offer a theoretical assessment of the two most important Coulombic excitations, 
AR and CM, in Si and Ge NCs. The Auger coefficients that we extracted can serve for the practical 
needs in the utilization of this process. For the efficiency enhancement via CM in Si and Ge NCs, 
the prospects look positive as the hot electrons receive most of the excess energy 
and they can undergo a CM within few picoseconds.

\begin{acknowledgments}
This work has been supported by the European FP6 Project SEMINANO with the 
Contract No. NMP4 CT2004 505285 and by the Turkish Scientific and Technical Council 
T\"UB\.ITAK with the Project No. 106T048. The computational resources are supplied in 
part by T\"{U}B\.ITAK through TR-Grid e-Infrastructure Project.
\end{acknowledgments}

\end{document}